\documentclass[prd,twocolumn,floatfix,preprintnumbers,showpacs]{revtex4}
\usepackage{graphicx}
\usepackage{epsfig}
\usepackage{amsfonts}

\newcommand{\lP}{\ell_{\rm P}}
\newcommand{\md}{{\rm d}}
\newcommand{\tr}{\mathop{\rm tr}}

\begin{document}

\preprint{IGPG--07/6--3}
\title{The radiation equation of state and loop quantum gravity corrections}
\author{Martin Bojowald}
\affiliation{Institute for Gravitational Physics and Geometry, The Pennsylvania State University,
104 Davey Lab, University park, PA 16802}\author{Rupam Das} 
\affiliation{Department of Physics and Astronomy, Vanderbilt University,
Nashville, TN 37235}
\affiliation{Institute for Gravitational Physics and Geometry, The Pennsylvania State University,
104 Davey Lab, University park, PA 16802}

\begin{abstract}
  The equation of state for radiation is derived in a canonical
  formulation of the electromagnetic field. This allows one to include
  correction terms expected from canonical quantum gravity and to
  infer implications to the universe evolution in radiation dominated
  epochs. Corrections implied by quantum geometry can be interpreted
  in physically appealing ways, relating to the conformal invariance of
  the classical equations.
\end{abstract}

\maketitle
\section{INTRODUCTION}
\label{sec:INTRODUCTION}

In theoretical cosmology, many insights can already be gained from
spatially isotropic Friedmann--Robertson--Walker models
\begin{equation}
 \md s^2= -\md\tau^2+a(\tau)^2\left(\frac{\md
r^2}{1-kr^2}+r^2(\md\vartheta^2+ \sin^2\vartheta\md\varphi^2)\right)
\end{equation}
with $k=0$ or $\pm 1$. The matter content in such a highly symmetric
space-time can only be of the form of a perfect fluid with
stress-energy tensor $T_{ab}=\rho u_au_b+P (g_{ab}+u_au_b)$ where
$\rho$ is the energy density of the fluid, $P$ its pressure and $u^a$
the 4-velocity vector field of isotropic co-moving observers. Once an
equation of state $P=P(\rho)$ is specified to characterize the matter
ingredients, the continuity equation $\dot{\rho}+3H(\rho+P)=0$ with
the Hubble parameter $H=\dot{a}/a$ allows one to determine the
behavior of $\rho(a)$ in which energy density changes during the
expansion or contraction of the universe. This function, in turn,
enters the Friedmann equation $H^2+k/a^2=8\pi G\rho/3$ and allows one
to derive solutions for $a(\tau)$.

In general, one would expect the equation of state $P=P(\rho)$ to be
non-linear which would make an explicit solution of the continuity and
Friedmann equations difficult. It is thus quite fortunate that in many
cases linear equations of state $P=w\rho$ with $w$ constant are
sufficient to describe the main matter contributions encountered in
cosmology at least phenomenologically. The influence of compact
objects on cosmological scales is, for instance, described well by the
simple dust equation of state $P(\rho)=0$. Relativistic matter, mainly
electromagnetic radiation, satisfies the linear equation of state
$P=\frac{1}{3}\rho$. The latter example is an exact equation
describing the Maxwell field, rather than an approximation for large
scale cosmology. It is thus, at first sight, rather surprising that
the dynamics of electromagnetic waves in a universe can be summarized
in such a simple equation of state irrespective of details of the
field configuration. The result follows in the standard way from the
trace-freedom of the electromagnetic stress-energy tensor and is thus
related to the conformal symmetry of Maxwell's equations. That the
availability of such a simple equation of state is very special for a
matter field can be seen by taking the example of a scalar field
$\phi$ with potential $V(\phi)$. In this case, we have an energy
density $\rho=\frac{1}{2}\dot{\phi}^2+V(\phi)$ and pressure
$P=\frac{1}{2}\dot{\phi}^2-V(\phi)$. Unless the scalar is free and
massless, $V(\phi)=0$ for which we have a stiff fluid $P=\rho$, there
is no simple relation between pressure and energy density
independently of a specific solution.

Any conformal symmetry such as that of elecromagnetism might be broken
by quantum effects especially when quantum gravity with its new scale
provided by the Planck length is taken into account. The coupling of
the electromagnetic field to geometry will then change, and exact
conformal symmetries can easily be violated. Accordingly, one expects
corrections from quantum gravity to the radiation equation of state
and corresponding effects in the universe evolution during radiation
dominated epochs. In loop quantum cosmology \cite{LivRev} equations of
state of matter fields are in general modified by perturbative
corrections at large scales and non-perturbative ones on small scales
\cite{InvScale}. This has mainly been studied so far for a scalar
field for which quantum modifications can be so strong that negative
pressure results independently of the chosen potential
\cite{Inflation}. The main reason is the fact that the isotropic
scalar field Hamiltonian $H_{\phi}=\frac{1}{2}a^{-3}p_{\phi}^2+
a^3V(\phi)$, where $p_{\phi}$ is the momentum of $\phi$, contains an
inverse power of the scale factor $a$. For quantum gravity, this
factor has to be quantized, too. Using the methods of \cite{QSDV}, it
turns out that inverse powers receive strong loop quantum corrections
at small length scales \cite{InvScale}. Accordingly, such
modifications play a role for effective equations describing the
universe after the big bang (or even during the quantum transition
through the big bang singularity). During later stages, modifications
are expected to decrease in size, but they might still be relevant due
to sometimes tight constraints on evolution parameters.

An extension to the usual matter ingredients of cosmology with linear
equations of state is, however, difficult since the modification is
based on quantizations of the fundamental field Hamiltonians.
Equations of state are obtained from fundamental Hamiltonians after an
analysis of the matter field equations, which can be difficult in
general especially when quantum effects are taken into account. The
only exception is the dust case since it implies a constant
Hamiltonian (the total mass of dust) which is straightforwardly
quantized without any corrections. Thus, although the dust energy
density is proportional to $a^{-3}$ and metric dependent in a way
which involves the inverse, it does not receive any modification since
the Hamiltonian, i.e.\ total energy $a^3\rho$, is the essential object
to be quantized. For radiation with $\rho\propto a^{-4}$ the
expectation is not clear since the total energy does behave like an
inverse power of $a$, but this follows only after an indirect analysis
of the field dynamics. It is not the solution $\rho(a)\propto a^{-4}$
of the continuity equation which is quantized but the original field
Hamiltonian from which the equation of state has to be derived first.
One thus has to go back to the fundamental Maxwell Hamiltonian, derive
energy density and pressure and see how quantum effects change the
equation of state. If this is completed, one may attempt to solve the
continuity equation to obtain corrections to $\rho(a)$.

We will derive such corrections in this article, using the canonical
quantization given by loop quantum gravity \cite{Rov,ALRev,ThomasRev}.
Candidates for Hamiltonian operators of the Maxwell field have been
proposed \cite{QSDV} which show several sources of correction terms.
To derive corrections to the equation of state, however, we need to
perform the usual calculation in a Hamiltonian formulation.  Thus, we
first present the canonical formulation for the free classical Maxwell
field to rederive the standard result for the equation of state
parameter $w$ without reference to an action or the stress-energy
tensor. Appropriate modifications to the matter Hamiltonian $H_{M}$
are then made to derive possible loop quantum gravity corrections to
the equation of state $w$. We will show that one case of corrections
results again in a linear equation of state, albeit in a corrected way
which depends on the basic discreteness scale of quantum gravity. In
this case we are able to express, as in the classical case, the full
field dynamics in terms of a simple modified $w$, and to solve
explicitly for $\rho(a)$. Our derivation takes into account
inhomogeneous field configurations and presents the first modified
equation of state obtained for a realistic matter source in loop
quantum gravity.

\section{Canonical Formulation}
\label{sec:canonical Formulation}

In a canonical formulation, the Hamiltonian $H_{M}$ rather than the
action is used to determine equations of motion of any function $f$ on
the phase space by means of Poisson brackets,
$\dot{f}=\left\{f,H_M\right\}$. The Poisson structure defines the
kinematical arena which follows from the field variables and momenta.
The basic configuration variable in a Lagrangian formulation of
Maxwell's field theory is the vector potential $A_{a}$ which
determines the field strength tensor
\begin{equation}
\label{Fieldtensor}
F_{ab}=\nabla_{a}A_{b} - \nabla_{b} A_{a}\,,
\end{equation}
where $\nabla_{a}$ is the covariant derivative operator. Notice that
$\nabla_{a}$ can be replaced by the partial derivative operator
$\partial_{a}$ even on a curved space-time since the field strength
tensor $F_{ab}$ is antisymmetric. The action for the free Maxwell
field in an arbitrary background space is given by
\begin{eqnarray}
\label{action}	
S_{M} &=& -\frac{1}{16\pi}\int \md^{4}x\sqrt{-g} F_{ab}F^{ab}\nonumber\\
&=& -\frac{1}{16\pi} \int \md^{4}x \sqrt{-g} F_{ab}F_{cd}g^{ac}g^{bd} 
\end{eqnarray}
where $g$ is the determinant of the Lorentzian space-time metric
$g_{ab}$.  From the action one obtains Maxwell's equations as the
Euler--Lagrange equations extremizing $S_M$.

A canonical formalism (Hamiltonian framework) is achieved by
performing a Legendre transform of this action $S_{M}$, replacing time
derivatives of configuration variables by momenta. This, as always,
requires one to treat space and time differently and is the reason why
the canonical formulation is not manifestly covariant. We introduce a
foliation of the space-time $\left(M,g_{ab}\right)$ by a family of
space-like hypersurfaces $\Sigma_t\colon t={\rm constant}$ in terms of
a time function $t$ on $M$. Canonical variables will depend on which
time function one chooses, but the resulting dynamics of observable
quantities will remain covariant. Furthermore, let $t^{a}$ be a
timelike vector field whose integral curves intersect each leaf
$\Sigma_t$ of the foliation precisely once and which is normalized
such that $t^{a}\nabla_{a}t=1$. This $t^{a}$ is the `evolution vector
field' along whose orbits different points on all
$\Sigma_t\equiv\Sigma$ can be identified. This allows us to write all
space-time fields in terms of $t$-dependent components defined on a
spatial manifold $\Sigma$. Lie derivatives of space-time fields along
$t^{a}$ are identified with `time-derivatives' of the spatial fields.

\subsection{Hamiltonian}

\begin{figure}
\begin{center}
\includegraphics[width=6cm]{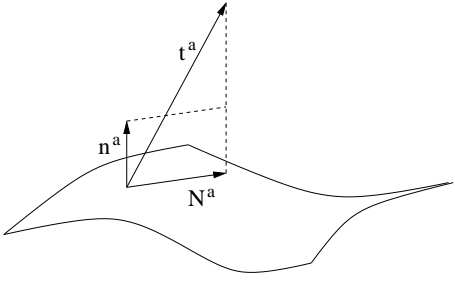}
\caption{Decomposition of the evolution vector field $t^a$ in terms of
the normal $n^a$ to spatial slices and a spacelike part $N^a$. \label{Split}}
\end{center}
\end{figure}

Let us, as illustrated in Fig.~\ref{Split}, decompose $t^{a}$ into
normal and tangential parts with respect to $\Sigma_{t}$ by defining
the lapse function $N$ and the shift vector $N^{a}$ as $t^{a}= Nn^{a}+
N^{a}$ with $N^{a}n_{a}=0$, where $n^{a}$ is the unit normal vector
field to the hypersurfaces $\Sigma_{t}$. The space-time metric
$g_{ab}$ induces a spatial metric $q_{ab}$ by the formula
$g_{ab}=q_{ab}-n_{a}n_{b}$. Now using $n^{a}=N^{-1}(t^{a}-N^{a})$ and
$q^{ab}=g^{ab}+n^{a}n^{b}$ to project fields normal and tangential to
$\Sigma_{t}$, we can decompose the field strength tensor $F_{ab}$ and
the action $S_{M}$ as follows:
\begin{widetext}
\begin{eqnarray}
\label{Fabdecomposition}
F_{ab}n^{a} &=& \frac {1}{N} \left(F_{ab}t^{a}-N^{a}F_{ab}\right)= \frac
{1}{N}
\left(\dot{A}_{b}-\partial_{b}\left(A_{a}t^{a}\right)-N^{a}F_{ab}\right)\,,\\
F_{ab}F^{ab} &=& F_{ab} F_{cd} g^{ac}g^{bd} = 
F_{ab} F_{cd} \left(q^{ac}-n^{a}n^{c}\right)\left(q^{bd}-n^{b}n^{d}\right) 
= F_{ab}F_{cd} q^{ac}q^{bd} - 2F_{ab} F_{cd}n^{a}n^{c}q^{bd}\nonumber \\ 
& =& F_{ab}F_{cd} q^{ac}q^{bd} -  \frac {2}{N^{2}} 
\left(\dot{A}_{b}-\partial_{b}\left(A_{a}t^{a}\right)-N^{a}F_{ab}\right)
\left(\dot{A}_{d}-\partial_{d}\left(A_{a}t^{a}\right)-
N^{c}F_{cd}\right)q^{bd}\,,
\label{FabFabdecomposition}
\end{eqnarray}
where $\dot{A}_{b}= {\cal L}_{t}A_{b}=
t^{a}\partial_{a}A_{b}+A_{a}\partial_{b}t^{a}$, and the action takes
the form
\begin{eqnarray}
\label{actiondecomposition}
S_{M} &=& -\frac{1}{16\pi}\int \md^{4}x \sqrt{-g} F_{ab}F^{ab} =
-\frac{1}{16\pi}\int \md t \int_{\Sigma_{t}}\md^{3}x N\sqrt{q} F_{ab}F^{ab}
\nonumber\ \\ &=& -\frac{1}{16\pi}\int \md t \int_{\Sigma_{t}}\md^{3}x
N\sqrt{q}\left( -\frac {2}{N^{2}}
\left(\dot{A}_{b}-\partial_{b}\left(A_{a}t^{a}\right)-N^{a}F_{ab}\right)
 \left(\dot{A}_{d}-\partial_{d}\left(A_{a}t^{a}\right)-
 N^{c}F_{cd}\right)q^{bd}+F_{ab}F_{cd} q^{ac}q^{bd}\right)\nonumber\
 \\ &=&\int \md t \int_{\Sigma_{t}}\md^{3}x \left(\frac{\sqrt{q}}{8\pi N}
\left(\dot{A}_{b}-\partial_{b}\left(A_{a}t^{a}\right)-N^{a}F_{ab}\right)
 \left(\dot{A}_{d}-\partial_{d}\left(A_{a}t^{a}\right)-
 N^{c}F_{cd}\right)q^{bd}-\frac{N \sqrt{q}}{16\pi}F_{ab}F_{cd}
 q^{ac}q^{bd}\right).
\end{eqnarray}
\end{widetext}
It follows that the conjugate momentum $\pi^{a}$ to the configuration
variable $A_{a}$ is given by
\begin{eqnarray}
\label{momentum}
 \pi^{e}=\frac{\delta S_{M}}{\delta \dot{A}_{e}} =
 \frac{\sqrt{q}}{4\pi
 N}\left(\dot{A}_{d}-\partial_{d}\left(A_{a}t^{a}\right)-
 N^{c}F_{cd}\right)q^{ed},
\end{eqnarray}
which is a densitized vector field because of the presence of
$\sqrt{q}$. Its physical interpretation is as the electric field
measured by an observer with 4-velocity $n^a$.  Now the action can be
expressed in terms of the canonical variables $A_{a}$ and $\pi^{a}$,
\begin{widetext}
\begin{eqnarray}
\label{actionpia}
S_{M}\left(A_{a},\pi^{a}\right)= \int \md t \int_{\Sigma_{t}}\md^{3}x
\left(\frac{2\pi N}{\sqrt{q}}\pi^{a}\pi^{b}q_{ab} - 
\frac{N \sqrt{q}}{16\pi}F_{ab}F_{cd} q^{ac}q^{bd}\right).
\end{eqnarray}
We can cast the action in equation (\ref{actionpia}) into the desired
form $S_{M}=\int
\md t\left[\int_{\Sigma_{t}}\md^{3}x\pi^{a}\dot{A}_{a} -
H_{M}\right]$ by writing the integrand in the following manner:
\begin{eqnarray}
\label{desiredaction}
S_{M}\left(A_{a},\pi^{a}\right) &=& \int \md t
\int_{\Sigma_{t}}\md^{3}x\left[\frac{4\pi N
}{\sqrt{q}}\pi^{a}\pi^{b}q_{ab}-N\left(\frac{2\pi
}{\sqrt{q}}\pi^{a}\pi^{b}q_{ab} + \frac{
\sqrt{q}}{16\pi}F_{ab}F_{cd} q^{ac}q^{bd}\right)\right] \nonumber\ \\
&=& \int \md t \int_{\Sigma_{t}}\md^{3}x\left[\pi^{a}
\left(\dot{A}_{a}-\partial_{a}\left(A_{d}t^{d}\right)-
N^{c}F_{ca}\right)-N\left(\frac{2\pi
}{\sqrt{q}}\pi^{a}\pi^{b}q_{ab} + \frac{
\sqrt{q}}{16\pi}F_{ab}F_{cd} q^{ac}q^{bd}\right)\right]\nonumber\ \\ 
&=& \int \md t
\int_{\Sigma_{t}}\md^{3}x\left[\pi^{a}\dot{A}_{a}+
 \left(A_{d}t^{d}\right)\partial_{a}\pi^{a}-N^{c}
 \pi^{a}F_{ca}-N\left(\frac{2\pi
 }{\sqrt{q}}\pi^{a}\pi^{b}q_{ab} + \frac{
\sqrt{q}}{16\pi}F_{ab}F_{cd} q^{ac}q^{bd}\right)\right]
\end{eqnarray}
\end{widetext}
having integrated by parts in the second term. This completes the
Legendre transform and we can read off the equations of motion from
equation (\ref{desiredaction}). First, since the momentum conjugate to
the time component of $A_{a}$ is absent, extremization of the action
with respect to $A_{a}t^{a}$ results in
\begin{equation}
\label {Gaussconstraint}
G = \partial_{a}\pi^{a}= 0
\end{equation}
as the usual Gauss constraint. The total Hamiltonian of the Maxwell
field then is
\begin{widetext}
\begin{equation}
\label{Hamiltonian}
H_{M} =
\int_{\Sigma_{t}}d^{3}x\left[-\left(A_{d}t^{d}\right)
 \partial_{a}\pi^{a}+N^{c}
 \pi^{a}F_{ca}+N\left(\frac{2\pi
 }{\sqrt{q}}\pi^{a}\pi^{b}q_{ab} + \frac{
\sqrt{q}}{16\pi}F_{ab}F_{cd} q^{ac}q^{bd}\right)\right]
\end{equation}
\end{widetext}
with two contributions
\begin{equation}
 D_c = \pi^{a}F_{ca}
\end{equation}
and
\begin{equation}
\label{hamiltonianconstraint}
{\cal H} = \frac{2\pi }{\sqrt{q}}\pi^{a}\pi^{b}q_{ab}
+ \frac{ \sqrt{q}}{16\pi}F_{ab}F_{cd} q^{ac}q^{bd}
\end{equation}
which, when added to the gravitational Hamiltonian, give matter
contributions to the diffeomorphism and Hamiltonian constraint,
respectively. From (\ref{hamiltonianconstraint}) we obtain the usual
expression $\int\md^3x{\cal H}$ for the energy of the electromagnetic
field.

\subsection{Equation of state}

The evolution equations can be obtained by evaluating the Poisson
brackets of $A_{a}$ and $\pi^{a}$ with the Hamiltonian. Although we
will not need the explicit form of these equations, we present them in
Appendix \ref{app:EOM} for the sake of completeness.  Here we rather
determine energy and pressure from our canonical expressions (see also
\cite{HamPerturb}) in order to formulate the equation of state.  The
matter Hamiltonian is directly related to energy density
\footnote{This is the usual term for energy per volume, and does not
  mean that $\rho$ is a geometrical density.} by
\begin{equation}
\label{rho}
\rho = \frac{1}{\sqrt{q}}\frac{\delta H_{M}}{\delta N},
\end{equation}
and thus, from equation (\ref{Hamiltonian}), it is
\begin{equation}
\label{rhom}
 \rho = \frac{2\pi
 }{{q}}\pi^{a}\pi^{b}q_{ab} +
 \frac{1}{16\pi}F_{ab}F_{cd} q^{ac}q^{bd}.
\end{equation}
The canonical formula for pressure is given by
\begin{equation}
\label{pressure}
P=-{\frac{2}{3N\sqrt{q}}}q_{ab}
{\frac{\delta{H_{M}}}{\delta{q_{ab}}}}=
{\frac{2}{3N\sqrt{q}}}q^{ab}{\frac{\delta{H_{M}}}{\delta{q^{ab}}}}
\end{equation}
as shown in Appendix~\ref{app:Pressure}. This gives
\begin{widetext}
\begin{eqnarray}
P&=& {\frac{2}{3N\sqrt{q}}}{q^{ef}}\left( {\frac{\pi
 N}{\sqrt{q}}}\pi^{a}
\pi^{b}\left(q_{ab}q_{ef}- 2q_{ae}q_{bf}\right)+
{\frac{\sqrt{q}N}{8\pi}}q^{ac}F_{ae}F_{cf}-
{\frac{\sqrt{q}N}{32\pi}}F_{ab}F^{ab}q_{ef}\right)\nonumber\\
&=&
{\frac{2}{3N\sqrt{q}}}\left({\frac{\pi
N}{\sqrt{q}}}\pi^{a}\pi^{b}q_{ab}+
{\frac{\sqrt{q}N}{32\pi}}F_{ab}F^{ab}\right)
={\frac{1}{3}}\left[ {\frac{2\pi
}{q}}\pi^{a}\pi^{b}q_{ab}+
{\frac{1}{16\pi}}F_{ab}F^{ab}\right]\,.\label{pressure2} 
\end{eqnarray}
\end{widetext}
Finally, the equation of state can easily be obtained from (\ref{rhom})
and (\ref{pressure2}):
\begin{equation}
\label{wr}
w=\frac{P}{\rho}=\frac{1}{3}
\end{equation}
which is the standard result.

\section{Quantization}
\label{sec:MODIFICATIONS}

Being interested in effects from quantum gravity, we have to quantize
metric components in the matter Hamiltonian (\ref{Hamiltonian}), not
just the matter field itself. Metric factors are thus not treated as a
given classical background but become operators. This requires a
quantum representation, which can only be found if we also use
canonical variables for the geometry. We thus need to use momenta of
$q_{ab}$ even though they do not appear in the matter Hamiltonian.

In loop quantum gravity, the basic objects appropriate for a canonical
quantization are constructed from a densitized triad $E^{a}_{i}$ and
the SU(2)-connection $A^{i}_{a}=\Gamma^{i}_{a}-\gamma K^{i}_{a}$ where
$\Gamma^{i}_{a}$ is the spin connection compatible with the triad,
$K^{i}_{a}$ the extrinsic curvature and $\gamma$ is the
Barbero-Immirzi parameter \cite{AshVar,AshVarReell,Immirzi}. Instead
of the spatial metric $q_{ab}$ we thus use the densitized vector
fields $E^a_i$ which are related to the metric by
$E^a_iE^b_i=q^{ab}\det q_{cd}$.

These fields cannot be quantized directly but must be integrated
suitably to remove local divergences in delta functions. The basic
ingredient of a loop quantization is to use holonomies $h_{e}(A)={\cal
  P}\exp\int_{e}A^{i}_{a}\dot{e}^{a}\tau_{i}\md t\in {\rm SU}(2)$ for
all curves $e\subset\Sigma$ and fluxes
$F_{S}(E)=\int_{S}E^{a}_{i}n_{a}\tau^{i}\md^{2}y$ for all surfaces
$S\subset\Sigma$ where $\tau_{i}$ are Pauli matrices, $\dot{e}^{a}$ is
the tangent vector to the edge $e$ and $n_{a}$ the co-normal to the
surface $S$. Thus the canonical quantization is performed by using
holonomies and fluxes as operators, turning their Poisson brackets
into commutators \cite{LoopRep,ALMMT}. A quantum representation is
easily constructed by using states which are functionals of
connections. Since holonomies are our basic connection dependent
operators, they serve to generate all states from a basic one which is
just a constant on the space of connections. All states are then
functionals depending on connections through holonomies, and they can
be associated with graphs collecting the edges of holonomies used in
the generation process. An orthonormal basis can be determined
explicitly in terms of spin network states \cite{RS:Spinnet}.

An immediate consequence of this quantization is that fluxes and
spatial geometrical operators such as area and volume
\cite{AreaVol,Area,Vol2} have discrete spectra containing zero.
Hence, their inverses do not exist as densely defined operators.
However, a quantization of the matter Hamiltonian such as
(\ref{Hamiltonian}) demands the quantization of such inverse
expressions since, e.g., $q^{-\frac{1}{2}}$ or the metric $q_{ab}$
which can only be obtained by inverting the densitized triad, appear
in the matter Hamiltonian. Therefore, the quantization of the matter
Hamiltonian seems, at first, to be seriously problematic. However, a
well defined quantization is possible after noticing that the Poisson
bracket of the volume with connection components,
\begin{equation} \label{poissonbracketofvolume}
 \left\{A^{i}_{a},\int\sqrt{\left|\det E \right|}d^{3}x\right\}=2\pi\gamma
 G \epsilon^{ijk} \epsilon_{abc} \frac{E^{b}_{j}
 E^{c}_{k}}{\sqrt{\left|\det E\right|}}=4\pi\gamma G e_a^i\,,
\end{equation}
amounts to an inverse of densitized triad components \cite{QSDI}. This
is written here in terms of the co-triad $e_a^i$ from which we can
directly obtain the metric $q_{ab}=e^i_ae^i_b$. Similar expressions
allow one to include the inverse determinant of the metric as we need
it in the Maxwell Hamiltonian. The left hand side of
(\ref{poissonbracketofvolume}) does not refer to inverse densitized
triad components and can be quantized: we can express the connection
component through holonomies, use the volume operator and turn the
Poisson bracket into a commutator. This observation enables us to
quantize inverse powers of the densitized triad. Leading to
well-defined operators, this quantization process implies
characteristic modifications of the classical expressions such as
(\ref{Hamiltonian}) on small scales, where densitized triad components
are small.  Moreover, since there are many different but classically
equivalent ways to rewrite expressions like
(\ref{poissonbracketofvolume}) for which the quantization would give
different results, there are quantization ambiguities. However,
several characteristic effects occur for any quantization choice such
that they can be studied reliably with phenomenological applications
in mind.

\subsection{Effective Maxwell Hamiltonian}

Hamiltonian operators of a quantum theory can, in semiclassical
regimes, be approximated by effective expressions which amend the
classical ones by quantum correction terms. The general procedure,
detailed in \cite{EffAc,Karpacz}, requires one to evaluate expectation
values of the Hamiltonian in suitable semiclassical states. A crucial
ingredient in loop quantum gravity is the discrete, non-local nature
of states written in terms of holonomies as basic objects. Although
Hamiltonian operators on such discrete lattice states are quite
complicated, expectation values can often be evaluated explicitly in
perturbative regimes where one assumes the geometry to be close to a
symmetric one. This is certainly allowed in our applications to derive
the effective equation of state of radiation in a flat FRW universe.
The background symmetry implies the existence of three approximate
spatial Killing vector fields $X^a_I$ generating transitive
isometries. We will only make use of this translational symmetry, not
of the additional rotations in the construction of states.  These
vector fields can be used as a tangent space basis, thus denoting
tensor indices for components in this basis by capital letters
$I,J,\ldots$

The background symmetry also has implications for the selection of
states of the quantum theory. A general quantization has to consider
arbitrary states, but for effective equations one computes expectation
values only in states suitable for a semiclassical regime. For
perturbative inhomogeneities, one can restrict lattices as they occur
in general graphs to regular cubic ones and thus simplify geometrical
operators. This has been developed recently in \cite{QuantCorrPert}
for metric perturbations as well as for a scalar field, and we can
directly apply the same techniques to the Maxwell Hamiltonian. We
refer the reader to this paper for more details.

\subsection{Gravitational variables and lattice states}

In a perturbative regime around a spatially flat isotropic solution,
one can choose the canonical variables to be given by functions
$(\tilde{p}^I(x),\tilde{k}_J(x))$ which determine a densitized triad
by $E^a_i=\tilde{p}^{(i)}(x)\delta^a_i$ and extrinsic curvature by
$K_a^i=\tilde{k}_{(i)}(x)\delta_a^i$. Thus, one can diagonalize the
canonical variables compared to the general situation where all matrix
elements of $E^a_i$ and $K_a^i$ would be independent. As seen in many
symmetric models, this simplifies the calculations considerably: it
allows one to replace involved SU(2) calculations by much simpler U(1)
calculations \cite{CosmoI,SphSymm}. SU(2) matrices arise because loop
quantum gravity is based on holonomies $h_e={\cal P}\exp(\int_e\md
t\dot{e}^aA_a^i\tau_i)$ of a connection $A_a^i$ related to extrinsic
curvature. For unrestricted connections, holonomies can take any SU(2)
value, but a diagonalization implies that all quantities can be
reduced to a maximal Abelian subgroup U(1).  Matrix elements of
Hamiltonians and other operators can then be computed in explicit
form.

Using properties of the general loop representation mentioned before,
basic variables of the quantum theory are, for a chosen lattice, U(1)
elements $\eta_{v,I}$ attached to a lattice link $e_{v,I}$ starting at
a vertex $v$ and pointing in direction $X_I^a$, and their conjugate
fluxes $F_{v,I}$. The U(1) elements $\eta_{v,I}$ appear as matrix
elements in SU(2) holonomies $h_{v,I}={\rm Re}\eta_{v,I}+2\tau_I {\rm
Im}\eta_{v,I}$ along edges $e_{v,I}$. Following the construction of
the Hilbert space using holonomies as ``creation'' operators by acting
on a state which is constant on the space of connections, a general
state is a functional $|\ldots,\mu_{v,I},\ldots\rangle=\prod_{v,I}
\eta_{v,I}^{\mu_{v,I}}$.  Allowing all possible values of assignments
of integers $\mu_{v,I}\in{\mathbb Z}$ to the lattice edges $e_{v,I}$,
this defines an orthonormal basis of the Hilbert space. Basic
operators are represented as holonomies
\begin{equation}
 \hat{\eta}_{v,I}|\ldots,\mu_{v',J},\ldots\rangle =
|\ldots,\mu_{v,I}+1,\ldots\rangle
\end{equation}
for each pair $(v,I)$ where all labels other than $\mu_{v,I}$ remain
unchanged, and fluxes
\begin{equation} \label{FluxVert}
 \hat{{\cal F}}_{v,I} |\ldots,\mu_{v',J},\ldots\rangle=
 2\pi\gamma\lP^2(\mu_{v,I}+\mu_{v,-I})
|\ldots,\mu_{v',J},\ldots\rangle
\end{equation}
where $\lP=\sqrt{\hbar G}$ is the Planck length and a subscript $-I$
means that the edge preceding the vertex $v$ in the chosen orientation
is taken. These and the following constructions are explained in more
detail in \cite{QuantCorrPert}.

Effective equations are obtained by taking expectation values of the
Hamiltonian operator and computing a continuum approximation of the
result (similar to a derivative expansion in low energy effective
actions). The result is a local field theory which includes quantum
corrections. This is done by relating holonomies
\begin{equation}
 \eta_{v,I} = \exp(i\smallint_{e_{v,I}}\md t
 \gamma\tilde{k}_I/2)\approx \exp(i\ell_0 \gamma\tilde{k}_I(v+I/2)/2)
\end{equation}
to continuum fields $\tilde{k}_I$ through mid-point evaluation on the
edges $e_{v,I}$ (denoted by an argument $v+I/2$ of the fields), and
similarly for fluxes
\begin{equation} \label{ScalarFlux}
 F_{v,I}=
\int_{S_{v,I}} \tilde{p}^I(y)\md^2y\approx
\ell_0^2\tilde{p}^I(v+I/2)\,.
\end{equation}
Although the non-local basic objects do not allow us to define
continuum fields at all spatial points, in a slowly-varying field
approximation the mid-point evaluations are sufficient to define the
continuum fields by interpolation.  Here, $\ell_0$ is the coordinate
length of lattice links. It does not appear in the quantum theory
which only refers to states and their labels $\mu_{v,I}$. This is
independent of coordinates and only makes use of an abstract, labelled
graph. The parameter $\ell_0$ only enters in the continuum
approximation since it is classical fields which are integrated and
related to holonomies and fluxes. These continuum fields, or tensor
components $\tilde{p}^I$ and $\tilde{k}_I$, must depend on which
coordinates are chosen to represent them. For the situation given
here, the combinations $p^I:=\ell_0^2\tilde{p}^I$ and
$k_I:=\ell_0\tilde{k}_I$, as they appear in holonomies and fluxes
evaluated for slowly-varying fields, are coordinate independent.

A further operator we can immediately define is the volume
operator. Using the classical expression $V=\int\md^3x
\sqrt{|\tilde{p}^1\tilde{p}^2\tilde{p}^3|}\approx \sum_v\ell_0^3
 \sqrt{|\tilde{p}^1\tilde{p}^2\tilde{p}^3|}= \sum_v\sqrt{|p^1p^2p^3|}$
 we introduce the volume operator $\hat{V}=\sum_{v} \prod_{I=1}^3
 \sqrt{|\hat{\cal F}_{v,I}|}$ which, using (\ref{FluxVert}), has
 eigenvalues
\begin{equation}\label{V_action}
 V(\{\mu_{v,I}\})= \left(2\pi\gamma\lP^2\right)^{3/2} \sum_{v}
 \prod_{I=1}^3\sqrt{ |\mu_{v,I}+\mu_{v,-I}|}\,.
\end{equation}
This operator is not only interesting for geometrical purposes, but
also for making use of the identity (\ref{poissonbracketofvolume}) or,
more generally,
\begin{equation}
 \{A_a^i,V_v^r\}=4\pi\gamma G\:rV_v^{r-1}e_a^i
\end{equation}
which gives inverse powers of the densitized triad for any $0<r<2$
often appearing in matter Hamiltonians. When quantizing this
expression using holonomies, the volume operator and a commutator for
the Poisson bracket, we obtain
\begin{eqnarray} 
\widehat{V_v^{r-1}{e}_I^i} &=& \frac{-2}{8\pi i r\gamma \lP^2 \ell_0}
\sum_{\sigma \in \{\pm 1\}}\sigma\tr(\tau^ih_{v,\sigma I}[h_{v,\sigma
I}^{-1},\hat{V_v}^r])\nonumber\\
 &=& \frac{1}{2\ell_0} (\hat{B}_{v,I}^{(r)} -
\hat{B}_{v,-I}^{(r)}) \delta^i_{(I)} =:\frac{1}{\ell_0} \hat{C}_{v,I}^{(r)}
\label{BC}\,.
\end{eqnarray}
For symmetry, we use both edges $e_{v,I}$ and $e_{v,-I}$ touching the
vertex $v$ along direction $X_I^a$. The operator $\hat B_{v,I}^{(r)}$
is obtained by taking the trace in (\ref{BC}) and using $h_{v,I}={\rm
Re}\eta_{v,I}+2\tau_I {\rm Im}\eta_{v,I}$,
\begin{equation}\label{B_def}
\hat B_{v,I}^{(r)} := \frac{1}{4 \pi i\gamma G \hbar r}\left(s_{v,I}
\hat V_v^r c_{v,I} - c_{v,I}\hat V_v^r s_{v,I}\right)
\end{equation}
with
\[
 c_{v,I}=\frac{1}{2}(\eta_{v,I}+\eta_{v,I}^*) \quad \mbox{ and }\quad
s_{v,I}=\frac{1}{2i}(\eta_{v,I}-\eta_{v,I}^*)\,.
\]

Such expressions can be used for the electric field part of
(\ref{Hamiltonian}) where the metric factor to be quantized is
\[
 \frac{q_{ab}}{\ell_0\sqrt{q}}=
\frac{e_a^ie_b^i}{\ell_0\sqrt{q}}\approx \frac{\ell_0^2
e_a^ie_b^i}{V_v}
\]
in terms of the volume $V_v\approx\ell_0^3\sqrt{q(v)}$ of a lattice
site. This can then be quantized, using (\ref{BC}) with $r=1/2$, to
\begin{equation}
\widehat{\frac{q_{IJ}}{\ell_0\sqrt{q}}}=
\widehat{(\ell_0V_v^{-1/2}e_I^i)} \widehat{(\ell_0V_v^{-1/2}e_J^i)}
= \hat{C}_{v,I}^{(1/2)} \hat{C}_{v,J}^{(1/2)}
\,.
\end{equation}
Noticing that the momentum $\pi^a$ of the electromagnetic
field is quantized, just as the densitized triad, by a flux operator
$\Pi_{v,I}:=\int_{S_{v,I}}\md^2y n_a\pi^a\approx
\ell_0^2\pi^I(v)$, the whole electric field term can be
written as
\begin{eqnarray*}
 H_{\pi} &=& 2\pi\int\md^3x N(x) \frac{q_{ab}}{\sqrt{q}}
\pi^a \pi^b \approx 2\pi \sum_v N(v) \ell_0^3
\frac{q_{ab}}{\sqrt{q}} \pi^a\pi^b\\
  &=&2\pi \sum_{v,I,J}N(v)
\frac{q_{IJ}}{\ell_0\sqrt{q}} \Pi_{v,I}\Pi_{v,J}
\end{eqnarray*}
which is then quantized to
\begin{equation}
 \hat{H}_{\pi}=2\pi \sum_vN(v) \hat{C}_{v,I}^{(1/2)} \hat{C}_{v,J}^{(1/2)}
\hat{\Pi}_{v,I}\hat{\Pi}_{v,J}\,.
\end{equation}

For the magnetic field term in (\ref{Hamiltonian}), at first sight, a
different metric expression arises: $\sqrt{q}q^{ac}q^{bd}$ which also
involves inverse components when expressed in terms of the densitized
triad. The term appears different from the electric field term and
could thus be quantized differently. However, noting
\begin{eqnarray*}
 F_{ab}F_{cd}q^{ac}q^{bd} &=& B^eB^f\epsilon_{eab}\epsilon_{fcd}
q^{ac}q^{bd}\\
 &=& \epsilon_{eab} B^eB^f q_{fd}\epsilon^{abd} q^{-1}= 2
q^{-1} q_{ab}B^aB^b
\end{eqnarray*}
in terms of the magnetic field $B^a=\epsilon^{abc}F_{bc}$ shows that
the metric dependence is the same as in the electric part. We thus
expect the same metric operator and correspondingly the same quantum
gravity corrections in both terms, although different ones are
mathematically possible owing to quantization ambiguities. The
magnetic contribution to the Maxwell Hamiltonian then is
\begin{eqnarray*}
 H_{B} &=& \frac{1}{8\pi}\int\md^3x N(x) \frac{q_{ab}}{\sqrt{q}}
B^a B^b \approx \frac{1}{8\pi} \sum_v N(v) \ell_0^3
\frac{q_{ab}}{\sqrt{q}} B^aB^b\\
  &=&\frac{1}{8\pi} \sum_{v,I,J}N(v)
\frac{q_{IJ}}{\ell_0\sqrt{q}} B_{v,I}B_{v,J}
\end{eqnarray*}
with the magnetic flux $B_{v,I}:= \int_{S_{v,I}}\md^2y n_aB^a\approx
\ell_0^2B^I(v)$. Magnetic flux components $B_{v,I}$ are quantized
using U(1) holonomies of the electromagnetic vector potential along
closed loops transversal to the direction $I$:
\[
 \hat{B}_{v,I}=
 \frac{1}{4}\sum_{J,K} \sum_{\sigma_J,\sigma_K\in\{\pm 1\}} \sigma_J\sigma_K 
 \epsilon^{IJK}\lambda_{v,\sigma_JJ,\sigma_KK}\,.
\]
We use the symbol $\lambda$ to distinguish an electromagnetic holonomy
$\lambda$ from a gravitational one, $\eta$. The loop holonomy
$\lambda_{v,\pm J,\pm K}$ is then computed around an elementary
lattice loop starting in $v$ in direction $\pm X_J^a$ and returning to
$v$ along $\pm X_K^a$. Summing over $J$, $K$ and the two sign factors
$\sigma_J$ and $\sigma_K$ accounts for all four loops starting in $v$
transversally to $e_{v,I}$.  The resulting quantized magnetic part of
the Hamiltonian is
\begin{equation}
 \hat{H}_{\pi}=\frac{1}{8\pi} \sum_vN(v) \hat{C}_{v,I}^{(1/2)}
 \hat{C}_{v,J}^{(1/2)}
\hat{B}_{v,I}\hat{B}_{v,J}
\end{equation}
with the same gravitational operator $\hat{C}_{v,I}^{(1/2)}
\hat{C}_{v,J}^{(1/2)}$ as in the electric term. It is thus natural to
use the same quantum operators and corresponding corrections in both
terms, even though mathematically it is possible to quantize them
differently. This aspect will be used in the following calculations.

\begin{widetext}
\subsection{Effective Hamiltonian and equation of state}

As in \cite{QuantCorrPert} we can include effects of the quantization
of metric coefficients by inserting correction functions in the
classical Hamiltonian which follow, e.g., from the eigenvalues
\cite{QuantCorrPert}
\begin{equation}
 C_{v,I}^{(1/2)}(\{\mu_{v',I'}\})= 2(2\pi\gamma\ell_{\rm P}^2)^{-1/4}
|\mu_{v,J}+ \mu_{v,-J}|^{1/4} |\mu_{v,K}+ \mu_{v,-K}|^{1/4}
\left(|\mu_{v,K}+ \mu_{v,-K}+1|^{1/4} - |\mu_{v,K}+
\mu_{v,-K}-1|^{1/4} \right)
\end{equation}
(where indices $J$ and $K$ are defined such that
$\epsilon_{IJK}\not=0$) of operators $\hat{C}^{(1/2)}_{v,I}$. Although
for large $\mu_{v,I}$ these eigenvalues approach the function
\[
C_{v,I}^{(1/2)}(\{\mu_{v',I'}\}) C_{v,J}^{(1/2)}(\{\mu_{v',I'}\})
\sim (2\pi\gamma\ell_{\rm
P}^2)^{-1/2} \frac{\prod_{K=1}^3\sqrt{|\mu_{v,K}+\mu_{v,-K}|}}{
|\mu_{v,I}+\mu_{v,-I}||\mu_{v,J}+\mu_{v,-J}|}
\]
\end{widetext}
expected classically for $q_{IJ}/\sqrt{q}= \sqrt{|p^1p^2p^3|}/p^Ip^J$
with a densitized triad $E^a_i=p^{(i)}\delta^a_i$ and using the
relation (\ref{FluxVert}) between labels and flux components, they
differ for values of $\mu_{v,I}$ closer to one. This deviation can,
for an isotropic background, be captured in a single correction
function
\begin{equation}
 \alpha_{v,K}=  \frac{1}{3} \sum_{I}
C_{v,I}^{(1/2)}(\{\mu_{v',I'}\})^2
\cdot
 \frac{\sqrt{2\pi\gamma\ell_{\rm P}^2}
 (\mu_{v,I}+\mu_{v,-I})^2}{\prod_{J=1}^3\sqrt{|\mu_{v,J}+\mu_{v,-J}|}}
\end{equation}
which would equal one in the absence of quantum corrections. This is
indeed approached in the limit where all $\mu_{v,I}\gg 1$, but for any
finite values there are corrections. If all $\mu_{v,I}>1$ one can
directly check that corrections are positive, i.e.\ $\alpha_{v,K}>1$
in this regime.  Expressing the labels in terms of the densitized
triad through fluxes (\ref{FluxVert}) results in functionals
\begin{equation} \label{alpha}
 \alpha[p^I(v)]= \alpha_{v,K}(4\pi\gamma\ell_{\rm P}^2 \mu_{v,I})
\end{equation}
which enter effective Hamiltonians.  The general expression one can
expect is thus
\begin{equation}
\label{modifiedhamiltonian}
H_{{\rm eff}} =
\int_{\Sigma}\md^{3}xN\left[\alpha[q_{cd}]\frac{2\pi
}{\sqrt{q}}\pi^{a}\pi^{b}q_{ab} +
\beta[q_{cd}]\frac{ \sqrt{q}}{16\pi}F_{ab}F_{cd} q^{ac}q^{bd}\right]
\end{equation}
with two possibly different correction functions $\alpha$ and $\beta$
depending on the lattice values $\mu_{v,I}$. As shown before, the case
$\alpha=\beta$ is preferred, and we will see soon that this has
implications for the effective equation of state. (In \cite{QSDV} a
Hamiltonian operator was introduced which did not use the same
quantizations for metric coefficients in the electric and magnetic
parts, thus giving $\alpha\not=\beta$. A quantization as described
here, using the same quantization in both parts, was formulated in
\cite{QFTonCSTI}. Phenomenological implications of a quantization of
the latter type, concerning Lorentz invariance, are discussed in
\cite{MaxwellLorentzInv}.) There are other possible sources for
corrections, such as higher order powers and higher derivatives of the
electric and magnetic fields. But these terms would not be metric
dependent and are thus not crucial for the following arguments.

Now using (\ref{modifiedhamiltonian}), we get the modified expression
\begin{eqnarray}
 \frac{1}{N}q^{ab}\frac{\delta H_{M}}{\delta q^{ab}} &=&
 -\frac{q_{ab}}{N}\frac{\delta H_{M}}{\delta q_{ab}}
={\frac{\pi
 }{\sqrt{q}}}\pi^{c}\pi^{d}q_{cd}\left(\alpha
 + 2q^{ab}\delta\alpha/\delta
q^{ab}\right)\nonumber\\
&&+{\frac{\sqrt{q}}{32\pi}}F_{cd}F^{cd}\left(\beta
 + 2q^{ab}\delta\beta/\delta
q^{ab}\right)\,,\label{hderivative3}
\end{eqnarray} 
depending on $\alpha$ and $\beta$.  For a nearly isotropic background
geometry, for instance, $\alpha$ only depends on the determinant $q$
of the spatial metric and, from Appendix B, $q^{ab}\delta\alpha/\delta
q^{ab}= - 3q\md\alpha/\md q$, which we assume in what follows.

The modified energy density and pressure then are
\begin{eqnarray}
\label{modifiedrho}
 \rho_{\rm eff} &=& \frac{2\pi
 }{q}\pi^{a}\pi^{b}q_{ab}\alpha + \frac{
 1}{16\pi}F_{ab}F_{cd} q^{ac}q^{bd}\beta\\
3P_{\rm eff}&=&{\frac{2\pi }{q}}\pi^{a}\pi^{b}q_{ab}\left(\alpha -
6q\md\alpha/\md q\right)\nonumber \\
&&+{\frac{1}{16\pi}}F_{ab}F^{ab}\left(\beta -
6q\md\beta/\md q\right) \nonumber\\ 
&=&{\frac{2\pi
}{q}}\pi^{a}\pi^{b}q_{ab}\alpha\left(1- 6\frac{\md\log \alpha}{\md\log
q}\right)\nonumber\\
&&+{\frac{1}{16\pi}}F_{ab}F^{ab}\beta\left(1- 6\frac{\md
\log\beta}{\md\log q}\right)\,.\label{modifiedp}
\end{eqnarray} 
It follows easily from (\ref{hderivative3}), (\ref{modifiedrho}) and
(\ref{modifiedp}) that the classical behavior is reproduced for
$\alpha$ = $\beta$ = 1. Interestingly, for $\alpha$ = $\beta$, the
equation of state $w$ can easily be computed and is modified as
\begin{eqnarray}
\label{modifiedw}
w_{\rm eff} = \frac{1}{3} - 2\frac{\md\log\alpha}{\md\log q}\,.
\end{eqnarray}
This modification is independent of the specific matter dynamics as in
the classical case, and it results in an equation of state which is
linear in $\rho$, but depends on the geometrical scales (and the
Planck length) through $\alpha$.

\section{COSMOLOGICAL APPLICATIONS}
\label{sec:COSMOLOGICALAPPLICATIONS}
In an isotropic and homogeneous universe (FRW), it follows from the
FRW metric and Einstein's equation that the evolution of the
energy density is given by the continuity equation, i.e.,
\begin{equation}
\label{continuity}
\dot{\rho} + 3\frac{\dot{a}}{a}\left(\rho+P\right)= 0,
\end{equation}
where $a$ is the scale factor and the dot indicates a proper time
derivative. Using the definition of the equation of state and
eliminating the time derivative, this equation can be cast into the
following useful form:
\begin{equation}
\label{usefulrho}
 \frac{\md\log \rho (a)}{\md\log
 a} = - 3 \left(1 + w(a)\right).
\end{equation}
Here we have shown the dependence of the equation of state on the
scale factor explicitly. It can easily be shown that the solution to
the above equation is
\begin{equation}
\label{solutionrho}
\rho (a)= \rho_{0}
\exp\left[-3\int\left(1+w(a)\right)\md\log a\right],
\end{equation}
where $\rho_{0}$ is the integration constant. Now by inserting the
modified equation of state in the radiation era, (\ref{modifiedw})
with $q=a^6$, we obtain
\begin{equation}
\label{correctedrho}
\rho(a)= \rho_{0} \alpha(a) a^{-4}.
\end{equation}
Again, for $\alpha$ = 1, we retrieve the classical result
$\rho(a)\propto a^{-4}$. Therefore, loop quantum gravity
corrections induced by discreteness of the flux operator are reflected
even in the evolution of the FRW universe.

Although one can write $\alpha$ as a function of the scale factor for
perturbations around a flat isotropic model, it is important to note
that corrections are well-defined even though the scale factor can be
rescaled arbitrarily. One can express $\alpha$ as a function of $a$
only after coordinates have been specified, such that there is no
ambiguity in relating the scale appearing in $\alpha$ (such as the
Planck length) to $a$. More precisely, we first obtained
$\alpha[p^I(v)]$ in (\ref{alpha}) with lattice values for the fluxes
$p^I(v)=\ell_0^2\tilde{p}^I(v)$ which are coordinate independent while
$\tilde{p}^I$ would be rescaled just as the scale factor. The quantum
state, through its lattice building blocks, unambiguously determines
the magnitude of the elementary variables as they appear in
corrections. Under rescalings or other coordinate changes, both the
classical field $\tilde{p}^I$ and the coordinate form of the lattice
change in such a way that elementary fluxes remain unchanged.  In a
nearly isotropic context, for instance, one has $p^I(v)\approx
p=\ell_0^2a^2$ spatially constant which can be related to the Hubble
scale by $Np^{3/2}=H^{-3}$. Here, we use the number $N$ of lattice
sites of elementary area $p$ in the Hubble volume $H^{-3}$ as a
measure of how fine the lattice is. Inserting all this in correction
functions yields $\alpha(p)=\alpha(N^{-2/3}H^{-2})$ expressed purely
in terms of coordinate independent quantities. The function $N$ enters
as an additional ingredient to describe the microstructure of the
underlying quantum state. In a given solution including the time
dependence $H(t)$ of the Hubble scale as well as a function $N(t)$
describing the quantum state one could relate all this, in a secondary
step, to the scale factor $a(t)$. But since the scale factor is not
the primary argument of correction functions, there is no problem with
rescalings.  See also \cite{InhomLattice} for further clarifications
of this issue which was not clear in all the literature on purely
homogeneous models.

\section{DISCUSSIONS}
\label{sec:DISCUSSIONS}

We have derived here the equation of state of the Maxwell field in a
canonical form, including corrections expected from loop quantum
gravity. In the canonical derivation, the reason for a linear equation
of state, which is trace-freedom in the Lagrangean derivation, is the
fact that the same metric dependent factor $q_{ab}/\sqrt{q}$
multiplies both terms in the Hamiltonian. The Maxwell Hamiltonian is
thus simply rescaled if the metric is conformally transformed, which
explains the conformal invariance of Maxwell's equations. This is
special for the Maxwell field and different from, e.g., a scalar field
with a non-vanishing potential.

The same fact allows one to quantize the Hamiltonian in a way which
affects both the electric and magnetic term in the same way, at least
as far as the metric dependence is concerned. One then obtains a
single correction function $\alpha=\beta$ which only corrects the
metric dependence of the total scale of the Hamiltonian. In this
sense, conformal invariance is preserved even after quantization. (But
this would not be the case if a quantization is used which results in
$\alpha\not=\beta$.)

This preservation of the form of the Hamiltonian explains why we are
still able to derive an equation of state independently of the
specific field dynamics and that it remains linear. However, the
classical value $w=\frac{1}{3}$ is corrected due to quantum effects in
the space-time structure. This modification is also understandable
from a Lagrangean perspective, together with basic information from
the loop quantization. Employing trace freedom of the stress-energy
tensor to derive the equation of state, we have to use the inverse
metric in $g^{ab}T_{ab}$. But from loop quantum gravity we know that,
when quantized, not all components of the inverse metric agree with
inverse operators of the quantization. For the scale factor of an
isotropic metric, for instance, we have
$\widehat{a^{-1}}\not=\mbox{``$\hat{a}^{-1}$''}$ since the right hand
side is not even defined \cite{InvScale}. While the left hand side is
defined through identities such as (\ref{poissonbracketofvolume}), it
satisfies $\widehat{a^{-1}}\hat{a}\not=1$ and thus shows deviations
from the classical expectation $a^{-1}a=1$ on small scales which were
captured here in correction functions. As derived in detail, this
implies scale dependent modifications to the equation of state
parameter $w_{\rm eff}$.

The result can also be interpreted in more physical terms. The
classical behavior $\rho(a)\propto a^{-4}$ can be understood as a
combination of a dilution factor $a^{-3}$ and an additional redshift
factor $a^{-1}$ for radiation in an expanding universe. As we have
seen, this is corrected to $\alpha(a)a^{-4}$ where $\alpha(a)$
corrects the metric factor $q_{ab}/\sqrt{q}\sim a^{-1}\delta_{ab}$.
Since this is only a single inverse power of $a$ for an isotropic
solution, we can interpret the result as saying that only redshift
receives corrections due to quantum effects on electromagnetic
propagation. The dilution factor due to expansion is unmodified,
except that the background evolution $a(t)$ itself receives
corrections. This agrees with the result for dust, which is only
diluted and has an unmodified equation of state even after
quantization \footnote{But it disagrees with \cite{Metamorph} both for
  dust and radiation, where a direct quantization of energy densities
  exclusively for isotropic fields was attempted.}. Unlike dust, for
radiation one has to refer to the inhomogeneous field and its quantum
Hamiltonian to derive a reliable equation of state, as presented here.

\begin{acknowledgments}
 MB was supported by NSF grant PHY0554771.
\end{acknowledgments}

\begin{appendix}

\section{Equations of motion}
\label{app:EOM}

It is straightforward to derive the equations of motion for the
canonical variables $A_{a}$ and $\pi^{a}$ from the Poisson
brackets of each of these variables with the matter Hamiltonian
$H_{M}$. Then
\begin{widetext}
\begin{eqnarray}
\label{eomfora}
 \dot{A}_{a}=\left\{A_{a},H_{M}\right\}=\frac{\delta H_{M}}{\delta
 \pi^{a}}=\partial_{a}\left(A_{c}t^{c}\right) +
 N^{c}F_{ca} +\frac{4\pi N}{\sqrt{q}} \pi^{c}q_{ca},
\end{eqnarray}
and
\begin{eqnarray}
\label{eomforpi}
 \dot{\pi}^{a}= \left\{\pi^{a},H_{M}\right\}=
 -\frac{\delta H_{M}}{\delta A_{a}}=
 \partial_{c}\left(N^{c}\pi^{a}\right) -
 \partial_{d}\left(N^{a}\pi^{d}\right)-
4\partial_{c}\left(N\sqrt{q}F_{ef}q^{ec}q^{fa}\right).
\end{eqnarray}
The modified Hamiltonian gives rise to the following new set of
equations of motion:
\begin{eqnarray}
\label{eomforam}
 \dot{A}_{{a}}=\left\{A_{a},H_{\rm eff}\right\}=\frac{\delta
 H_{\rm eff}}{\delta \pi^{a}}=\partial_{a}\left(A_{c}t^{c}\right) +
 N^{c}F_{ca} +\frac{4\pi N}{\sqrt{q}}
 \alpha\left(q\right)\pi^{c}q_{ca},
\end{eqnarray}
and
\begin{eqnarray}
\label{eomforpim}
 \dot{\pi}^{a}= \left\{\pi^{a},H_{\rm eff}\right\}=
 -\frac{\delta H_{\rm eff}}{\delta A_{a}}=
 \partial_{c}\left(N^{c}\pi^{a}\right) -
 \partial_{d}\left(N^{a}\pi^{d}\right)-4\partial_{c}\left(N\beta\left(q
 \right)\sqrt{q}F_{ef}q^{ec}q^{fa}\right),
\end{eqnarray}
\end{widetext}
where $H_{\rm eff}$ is the effective Hamiltonian of the Maxwell's
field ($H_{M}$ with $\alpha$ and $\beta$ inserted).

\section{Pressure}
\label{app:Pressure}

The general, thermodynamical definition of pressure is the negative
change of energy by volume, which we can write as
\begin{equation}
 P=-\frac{1}{N}\frac{\delta H}{\delta \sqrt{q}}
\end{equation}
whenever the Hamiltonian $H=\int\md^3x N(x){\cal H}(x)$ is depends
isotropically on the metric. Otherwise, one has to use all components
of the stress tensor $\delta H/\delta q^{ab}$ which is not
proportional to the identity. The derivative by the determinant of the
metric can be expressed in terms of metric components by using a
suitable change of variables which includes $q$ as an independent one.
We thus introduce $q_{ab}=:q^{1/3} \bar{q}_{ab}$ with $\det
\bar{q}_{ab}=1$ such that $\partial q_{ab}/\partial q=
\frac{1}{3}q^{-1}q_{ab}$ where all components of $\bar{q}_{ab}$ are
kept fixed in the partial derivative. This is exactly what we need to
compute pressure since only the volume but not the shape of the fluid
is varied.
This change of variables implies
\[
 \frac{\delta}{\delta \sqrt{q}}=2\sqrt{q}
 \frac{\delta}{\delta q}
 =2\sqrt{q}\sum_{ab}\frac{\partial q_{ab}}{\partial q}
\frac{\delta}{\delta q_{ab}}=
\frac{2}{3\sqrt{q}}\sum_{ab} q_{ab}\frac{\delta}{\delta
q_{ab}}
\]
and thus
\begin{equation}
 P=-\frac{2}{3N\sqrt{q}} q_{ab}\frac{\delta H}{\delta q_{ab}}\,.
\end{equation}

We can also verify this by comparing the dynamical effects of $H$ on
the metric with the Raychaudhuri equation expressed in terms of the
canonical variables which for simplicity we do for homogeneous
metrics. Using the following definitions for the extrinsic curvature
tensor $K_{ab}=\nabla_{a}n_{b}$ (which turns out to be automatically
spatial and symmetric without projection if homogeneity is used), the
expansion parameter $\theta=K_{ab}q^{ab}$ and the shear
$\sigma_{ab}=K_{(ab)}-\frac{1}{3}
\theta q_{ab}$, the canonical momentum conjugate to $q_{ab}$ derived from the
gravitational Lagrangian is
\[
 \pi^{ab}=
 \frac{\sqrt{q}}{16\pi G}\left(K^{ab}-K^{c}_{c}q^{ab}\right)=
\frac{\sqrt{q}}{16\pi G}\left(\sigma^{ab}-\frac{2}{3}\theta
 q^{ab}\right)
\]
where $G$ is the gravitational constant.  Then the Raychaudhuri
equation in terms of the canonical variables takes the following form:
\begin{equation}
\label{thetadot}
	\dot{\theta} = -8\pi G \frac{\md}{\md t}\left(\frac{\pi^{ab}
	q_{ab}}{\sqrt{q}}\right)\,.
\end{equation}
The canonical equations of
motion, in the presence of a matter Hamiltonian $H$ added to the
gravitational Hamiltonian to form $H_{\rm Total}$, become
\begin{equation}
\label{qdot}
 \dot{q}_{ab} = \frac {\delta H_{\rm Total}} {\delta \pi^{ab}} =
 \frac{16\pi GN}{\sqrt{q}}\left(2\pi_{ab}-q_{ab}\pi^{c}_{c}\right)
 + 2D_{(a}N_{b)}
\end{equation}
and
\begin{widetext}
\begin{eqnarray}
\label{pidot}
 \dot{\pi}^{ab} &=& -\frac {\delta H_{\rm Total}} {\delta q_{ab}} =
-\frac{N\sqrt{q}}{16\pi G}\left(^{(3)}R^{ab} - 
{\frac{1}{2}} {^{(3)}R}q^{ab}\right) +
\frac{8\pi GN}{\sqrt{q}}q^{ab}\left(\pi_{cd}\pi^{cd}-\frac{1}{2}
\pi^{2}\right) -\frac{32\pi GN}{\sqrt{q}} q^{ab}\left(\pi^{ac}\pi_{c}^{b} -
\frac{1}{2} \pi \pi^{ab}\right) \nonumber\\ &&
- \frac{\delta H}{\delta q_{ab}}
 + \frac{\sqrt{q}}{16\pi G}\left(D^{a}D^{b}N - 
q^{ab}D^{c}D_{c}N\right)
+ \sqrt{q}D_{c}\left(\frac{N^{c}\pi^{ab}}{\sqrt{q}}\right) -
2\pi^{c(a}D_{c}N^{b)},
\end{eqnarray}
\end{widetext}
where $D_{a}$ is the derivative operator compatible with
$q_{ab}$. Variation of the total action with respect to the lapse
function $N$ yields the Hamiltonian constraint equation
\begin{equation}
\label{hconstraint}
	-\frac{\sqrt{q}}{16\pi G}{}^{(3)}R + \frac{16\pi G}{\sqrt q}\left(\pi^{ab}\pi_{ab}-\frac{1}{2}\pi^{2}\right)+
H=0.
\end{equation}
Upon inserting equations (\ref{qdot}), (\ref{pidot}), and
(\ref{hconstraint}) into equation (\ref{thetadot}), the Raychaudhuri
equation becomes
\begin{eqnarray}
\label{thetadot2}
\frac{\dot{\theta}}{N} &=& -{\frac {1}{3}}\theta^{2} -\sigma^{ab}\sigma_{ab} 
- {\frac{4\pi G}{N\sqrt{q}}}H+{\frac{8\pi
G}{N\sqrt{q}}}q_{ab}{\frac{\delta{H}}{\delta{q_{ab}}}}\\
&& + D^{a}D_{a}N -
8\pi G D_{c}\left(\frac{N^{c}\pi^{a}_{a}}{\sqrt{q}}\right) + \frac {16
\pi G}{\sqrt{q}} \pi^{ca}D_{c}N_{a}\,,\nonumber
\end{eqnarray}
which, for a homogeneous universe, reduces to 
\begin{eqnarray}
\label{thetadot3}
\frac{\dot{\theta}}{N} = -{\frac {1}{3}}\theta^{2} -\sigma^{ab}\sigma_{ab} 
- {\frac{4\pi G}{N\sqrt{q}}}H+{\frac{8\pi
G}{N\sqrt{q}}}q_{ab}{\frac{\delta{H}}{\delta{q_{ab}}}}.
\end{eqnarray}
On the other hand, for a perfect fluid distribution, the Raychaudhuri
equation is found to be
\begin{eqnarray}
\label{thetadot4}
\frac{\dot{\theta}}{N} = -{\frac {1}{3}}\theta^{2} -\sigma^{ab}\sigma_{ab} 
- {4\pi G}\left(\rho+3P\right).
\end{eqnarray}
Now comparing equation (\ref{thetadot3}) with equation
(\ref{thetadot4}), we verify the canonical formula for the average
pressure for a perfect fluid distribution in an anisotropic geometry
\begin{equation}
\label{pressure1}
P=-{\frac{2}{3N\sqrt{q}}}q_{ab}{\frac{\delta{H}}{\delta{q_{ab}}}}=
{\frac{2}{3N\sqrt{q}}}q^{ab}{\frac{\delta{H}}{\delta{q^{ab}}}}.
\end{equation}

\end{appendix}


\begin{thebibliography}{10}

\bibitem{LivRev}
M. Bojowald, Living Rev.\ Relativity {\bf 8},  11  (2005).

\bibitem{InvScale}
M. Bojowald, Phys.\ Rev.\ D {\bf 64},  084018  (2001).

\bibitem{Inflation}
M. Bojowald, Phys.\ Rev.\ Lett. {\bf 89},  261301  (2002).

\bibitem{QSDV}
T. Thiemann, Class.\ Quantum Grav. {\bf 15},  1281  (1998).

\bibitem{Rov}
C. Rovelli, {\em Quantum Gravity} (Cambridge University Press, Cambridge, UK,
  2004).

\bibitem{ALRev}
A. Ashtekar and J. Lewandowski, Class.\ Quantum Grav. {\bf 21},  R53  (2004).

\bibitem{ThomasRev}
T. Thiemann, gr-qc/0110034.

\bibitem{HamPerturb}
M. Bojowald {\it et~al.}, Phys.\ Rev.\ D {\bf 74},  123512  (2006).

\bibitem{AshVar}
A. Ashtekar, Phys.\ Rev.\ D {\bf 36},  1587  (1987).

\bibitem{AshVarReell}
J.~F. Barbero~G., Phys.\ Rev.\ D {\bf 51},  5507  (1995).

\bibitem{Immirzi}
G. Immirzi, Class.\ Quantum Grav. {\bf 14},  L177  (1997).

\bibitem{LoopRep}
C. Rovelli and L. Smolin, Nucl.\ Phys.\ B {\bf 331},  80  (1990).

\bibitem{ALMMT}
A. Ashtekar {\it et~al.}, J.\ Math.\ Phys. {\bf 36},  6456  (1995).

\bibitem{RS:Spinnet}
C. Rovelli and L. Smolin, Phys.\ Rev.\ D {\bf 52},  5743  (1995).

\bibitem{AreaVol}
C. Rovelli and L. Smolin, Nucl.\ Phys.\ B {\bf 442},  593  (1995), erratum:
  Nucl.\ Phys.\ B {\bf 456}, 753 (1995).

\bibitem{Area}
A. Ashtekar and J. Lewandowski, Class.\ Quantum Grav. {\bf 14},  A55  (1997).

\bibitem{Vol2}
A. Ashtekar and J. Lewandowski, Adv.\ Theor.\ Math.\ Phys. {\bf 1},  388
  (1997).

\bibitem{QSDI}
T. Thiemann, Class.\ Quantum Grav. {\bf 15},  839  (1998).

\bibitem{EffAc}
M. Bojowald and A. Skirzewski, Rev.\ Math.\ Phys. {\bf 18},  713  (2006).

\bibitem{Karpacz}
M. Bojowald and A. Skirzewski,  in {\em Current Mathematical Topics in
  Gravitation and Cosmology (42nd Karpacz Winter School of Theoretical
  Physics)}, Int.\ J.\ Geom.\ Meth.\ Mod.\ Phys.\ {\bf 4},  25--52
 (2007), hep-th/0606232.

\bibitem{QuantCorrPert}
M. Bojowald, H. Hern\'andez, M. Kagan, and A. Skirzewski, Phys.\ Rev.\ D
{\bf 75}, 064022 (2007).

\bibitem{CosmoI}
M. Bojowald, Class.\ Quantum Grav. {\bf 17},  1489  (2000).

\bibitem{SphSymm}
M. Bojowald, Class.\ Quantum Grav. {\bf 21},  3733  (2004).

\bibitem{QFTonCSTI}
H. Sahlmann and T. Thiemann, Class.\ Quantum Grav. {\bf 23},  867  (2006).

\bibitem{MaxwellLorentzInv}
C.~N. Kozameh and M.~F. Parisi, Class.\ Quantum Grav. {\bf 21},  2617  (2004).

\bibitem{InhomLattice}
M. Bojowald, Gen.\ Rel.\ Grav. {\bf 38},  1771  (2006).

\bibitem{Metamorph}
P. Singh, Class.\ Quantum Grav. {\bf 22},  4203  (2005).

\end{thebibliography}

\end{document}